\documentclass[conference]{IEEEtran}
\usepackage[english]{babel}
\usepackage{float}
\usepackage{amsmath}
\usepackage{graphicx}
\usepackage[colorlinks=true, allcolors=blue]{hyperref}
\usepackage{placeins} 
\usepackage{caption}
\title{PlatonSPAD: A novel SPAD sensor for large-scale high-resolution particle detectors}

\author{
    \IEEEauthorblockN{
        Kodai Kaneyasu\IEEEauthorrefmark{1}, 
        Till Dieminger\IEEEauthorrefmark{2}, 
        Matthew Franks\IEEEauthorrefmark{2}, 
        Davide Sgalaberna\IEEEauthorrefmark{2}, 
        Claudio Bruschini\IEEEauthorrefmark{1}, 
        Edoardo Charbon\IEEEauthorrefmark{1}
    }
    \IEEEauthorblockA{
        \IEEEauthorrefmark{1}Advanced Quantum Architecture Laboratory (AQUA), EPFL, Neuchâtel, Switzerland\\
        \IEEEauthorrefmark{2}Institute for Particle Physics and Astrophysics (IPA), ETH Zürich, Zürich, Switzerland\\
        Email: \{kodai.kaneyasu, claudio.bruschini, edoardo.charbon\}@epfl.ch, \\
        tilld@phys.ethz.ch,  mfranks@ethz.ch,  davide.sgalaberna@cern.ch 
    }
}

\begin{document}
\setlength{\abovedisplayskip}{10pt}
\setlength{\belowdisplayskip}{10pt}
\setlength{\abovedisplayshortskip}{0pt}
\setlength{\belowdisplayshortskip}{0pt}
\maketitle
\setlength{\textfloatsep}{7pt}
\begin{abstract}
High-resolution 3D tracking with sub-nanosecond timing is required for the detection of elementary particles, such as neutrinos. Conventional detectors, which utilize analog silicon photomultipliers, face challenges in balancing spatial resolution and scalability. To address this issue, a CMOS single-photon avalanche diode (SPAD)-based high-resolution particle detector is being developed. This work presents a study on SPAD layout optimization and a 4×4 SPAD macropixel module, fabricated in 110 nm CIS technology. Measurement results confirm that high-fill-factor designs improve photon detection efficiency without significant noise degradation. Furthermore, event-driven photon mapping and time stamping, enabled by time-to-digital converters and dedicated pixel circuits integrated into the 4×4 SPAD macropixel, were successfully demonstrated. This work is an essential step towards a sensor that detects probabilistic particle interactions and it lays the groundwork for the development of future large-scale SPAD-based particle detectors.
\end{abstract}

\section{Introduction}
In high-energy physics (HEP) experiments, particle detectors often require high-resolution 3D tracking capabilities with sub-nanosecond time-stamping precision. An example is neutrino detection at the J-PARC proton accelerator in Japan for the T2K experiment \cite{T2K2011} and, in the near future, the Hyper-Kamiokande experiment \cite{HyperK_Snowmass2022}. These experiments aim to study matter-antimatter asymmetry in neutrino flavor oscillations, which could help explain the lack of antimatter in the universe. A key challenge in this field is the limited modeling accuracy of nuclear processes involved in neutrino-nucleus interactions, making precise 3D imaging detectors essential.  

\begin{figure}[htbp]
    \centering
    \includegraphics[width=\linewidth]{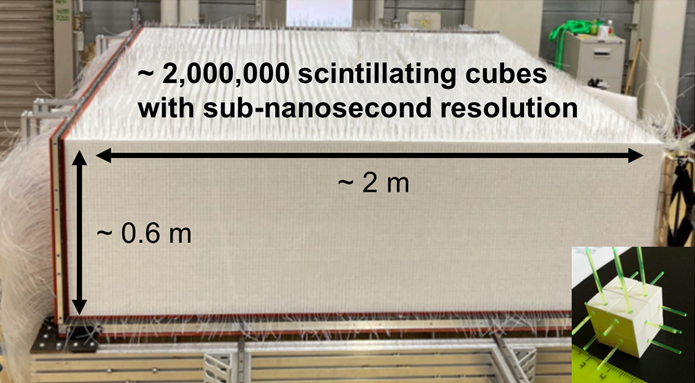}
    \caption{Picture of the SuperFGD detector showing its dimensions. \cite{Sgalaberna2024}\cite{Blondel2018}}
    \label{SuperFCD}
\end{figure}

Although organic scintillators have a nanosecond-level timing response, state-of-the-art neutrino detectors face a trade-off between achieving high spatial resolution (mm or sub-mm) and managing the prohibitive number of electronic readout channels \cite{Douqa2020, JesusValls2022}, which is driven by the ton-scale mass required for detection. Figure~\ref{SuperFCD} illustrates the near detector (ND) facility of the T2K experiment, known as SuperFGD, which consists of approximately 2,000,000 scintillator cubes with wavelength-shifting fibers coupled to analog silicon photomultipliers (SiPMs) at their ends. The spatial resolution of this system is fundamentally limited by the scintillator cube size, which is 1 cm³. To overcome this limitation, a paradigm shift toward integrated single-photon detection systems is required.  

\begin{figure}[htbp]
    \centering
    \includegraphics[width=\linewidth]{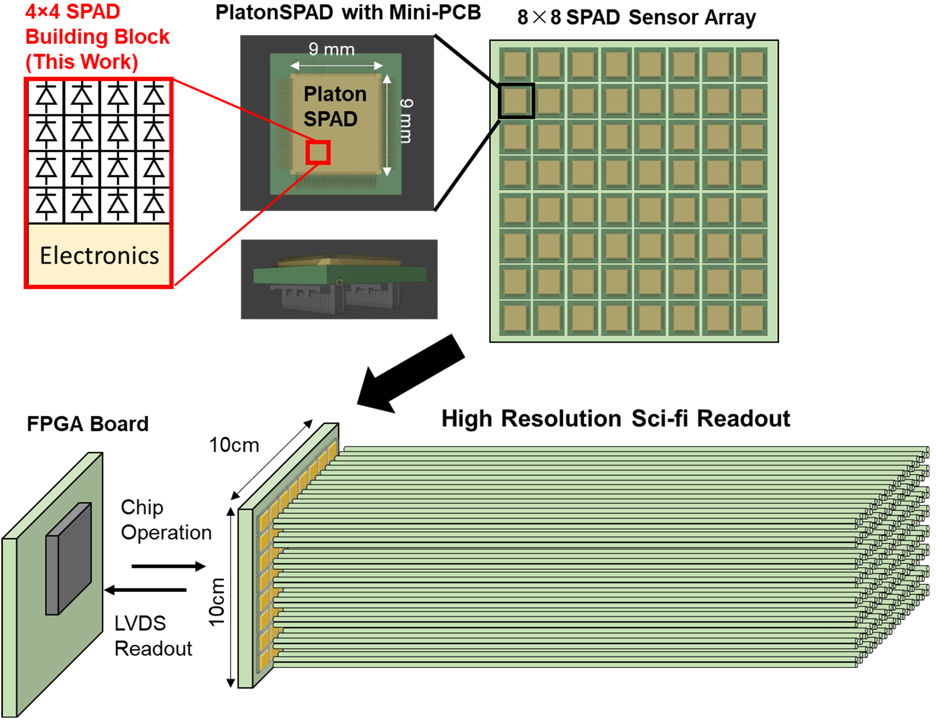}
    \caption{Concept of PlatonSPAD, the large-scale high-resolution particle tracking system.}
    \label{High_res}
\end{figure}

SPAD arrays and digital SiPMs (dSiPMs), with their CMOS compatibility and scalability, play a key role in enabling this transition \cite{Morimoto2020}. The feasibility of submillimeter-level charged particle tracking using scintillating fibers coupled with SPAD sensors has already been demonstrated \cite{Franks2024}, with prior studies highlighting the use of SPADs in other high-energy physics applications \cite{Ripiccini,Bruschini2014,Dolenec2023}. To further improve performance, a large-scale high-resolution particle detector is currently being designed, which incorporates a scintillating fiber bundle and an array of large-format SPADs (Figure~\ref{High_res}). This system requires a dedicated SPAD, optimized for this application, to be tiled into large arrays with minimal dead area.  

In this paper, we present a study on candidate SPAD layouts and on 4×4 SPAD macropixel modules fabricated in 110-nm CIS technology. These modules will serve as the foundation for the final large-format SPAD sensor. The few hundred picosecond time resolution of CMOS SPAD arrays enables noise rejection by signal coincidence, as well as precise time measurements of particle interactions, making them highly suitable for applications requiring accurate temporal resolution.

\FloatBarrier 

\section{Design Considerations}
The required specifications for the future high-resolution particle detector determine our SPAD layout optimization strategy and the architecture of 4×4 SPAD macropixels.

\subsection{SPAD Layout Study}
\begin{figure}[t]
    \centering
    \includegraphics[width=\linewidth]{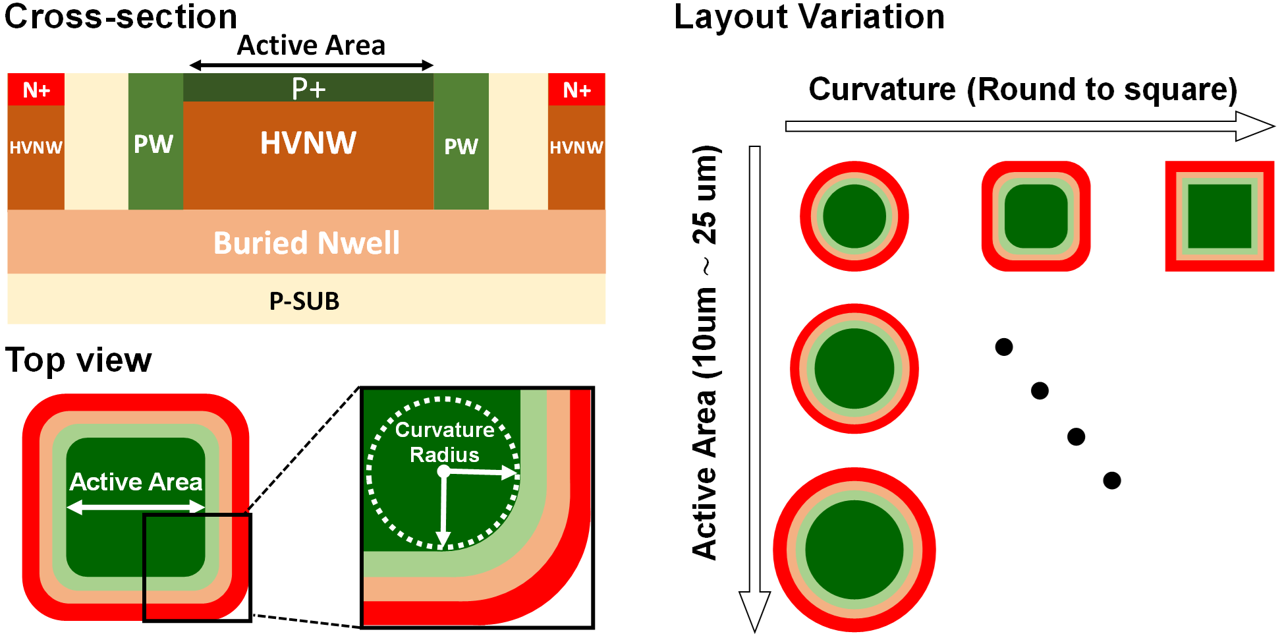}
    \caption{Cross-section and layout variations of the SPADs in 110 nm CIS Technology. \cite{Karaca2024}}
    \label{SPAD_layout}
\end{figure}

To detect sparse scintillation photons generated by charged particles, the SPAD sensor must achieve high photon detection efficiency (PDE), defined as:
\[
PDE = \mathit{PDP} \times \mathit{FF}
\]
where PDP is the photon detection probability and FF the fill factor. Meanwhile, minimizing dark count rate (DCR) is also crucial, as high DCR makes it more challenging to distinguish between sparse signals and background noise. This study explores two layout optimization approaches to balance PDE and potential performance trade-offs. The first approach increases the active area, which enhances PDE but may also lead to higher DCR. The second approach modifies the SPAD shape from round to square, reducing curvature but potentially introducing premature edge breakdown, which causes a severe degradation of PDP and an increase of DCR. It is important to note that a square-shaped SPAD is advantageous for FF, especially when SPAD arrays are implemented with well or guard-ring sharing \cite{Morimoto2020_HighFF}, as it has less dead area. The cross-sectional structure and layout variations are illustrated in Figure~\ref{SPAD_layout}.

\subsection{4\texttimes  4 SPAD macropixel architecture}
In application fields, where SPADs are commonly utilized, such as fluorescence lifetime imaging microscopy (FLIM)  and light detection and ranging (LiDAR), time resolution is typically achieved using either gating \cite{Ulku2019} or a time-to-digital converter (TDC) \cite{Ximenes2019}. In the previous particle tracking demonstration \cite{Franks2024}, the sensor operated with a gating scheme, where it was selectively activated within a predefined time window. This architecture is particularly beneficial in environments where photon arrival times are well-correlated with an external trigger. In contrast, however, particle interactions occur probabilistically within a time window of the order of seconds or, like at the T2K neutrino experiments, microseconds, making it difficult to apply a conventional gating architectures.

\begin{figure}[htbp]
    \centering
    \includegraphics[width=1\linewidth]{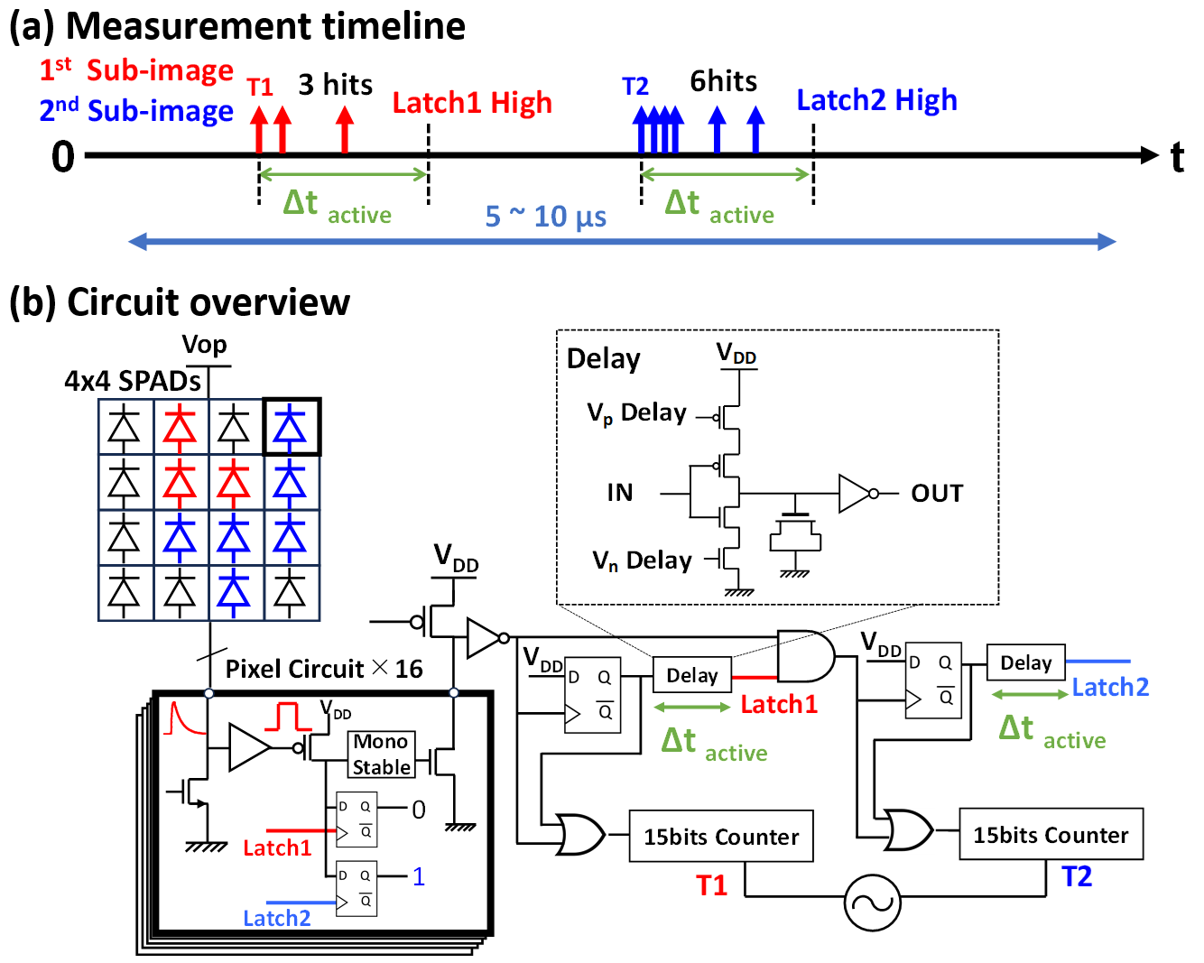}
    \caption{System overview of the designed 4×4 SPAD Sensor micropixel module. (a) Example of the measurement timeline in a neutrino detection experiment. (b) Circuit overview of the 4×4 SPAD Sensor micropixel module.}
    \label{System_Overview}
\end{figure}

To detect probabilistic events, where sub-nanosecond measurements of the particle interaction time are often needed, the designed sensor operates in an event-driven manner, storing pixel maps and time stamps with hundreds of picoseconds resolution accordingly. Figure~\ref{System_Overview} presents the system architecture of the macropixel module, which comprises SPADs, individual pixel circuits, and two TDCs with a 200-ps LSB and 15-bit memory, covering the entire time window of 5 $\mu$s . Upon SPAD activation, the sensor records the corresponding timestamp in the TDC while simultaneously triggering a tunable delay, implemented via a current-starved inverter (CSI), referred to as $\Delta t$ active. At the end of this interval, the states of the firing SPADs are latched by activating Latch1 in the pixel circuit. This sequence can be executed twice per measurement, increasing the probability of capturing the target signal. Finally, the sensor produces two sub-images, each associated with one timestamp each (first hit).

\begin{figure}[htbp]
    \centering
    \includegraphics[width=1\linewidth]{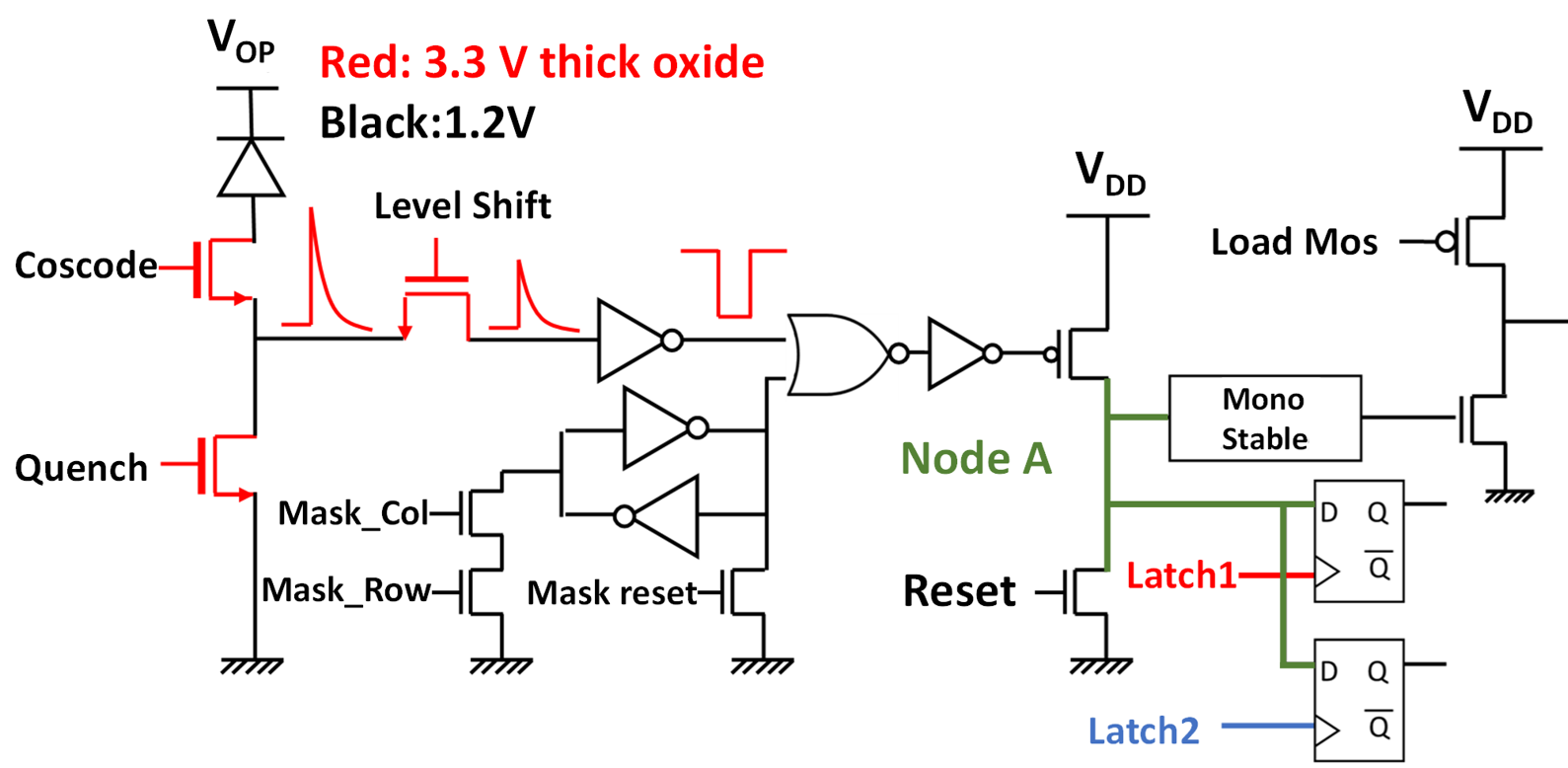}
    \caption{Pixel circuit of 4×4 SPAD Sensor micropixel module.}
    \label{pixel_Circuit}
\end{figure}

Figure~\ref{pixel_Circuit} presents the detailed schematic of the pixel circuit, which employs a passive recharge and passive quench architecture. A cascode transistor is connected in series with the quench transistor to increase the maximum excess bias of the SPADs. An SRAM module is integrated to mask noisy pixels. When a SPAD is triggered, the corresponding pixel temporarily stores a binary value at Node A while a monostable circuit generates a pulse signal to trigger the TDC. After Latch1 or Latch2 is activated, the value is transferred to a D flip-flop (DFF), where it is securely stored.


\section{Characterization}
\subsection{Single SPAD measurement}
\begin{figure}[t]
    \centering
    \includegraphics[width=\linewidth]{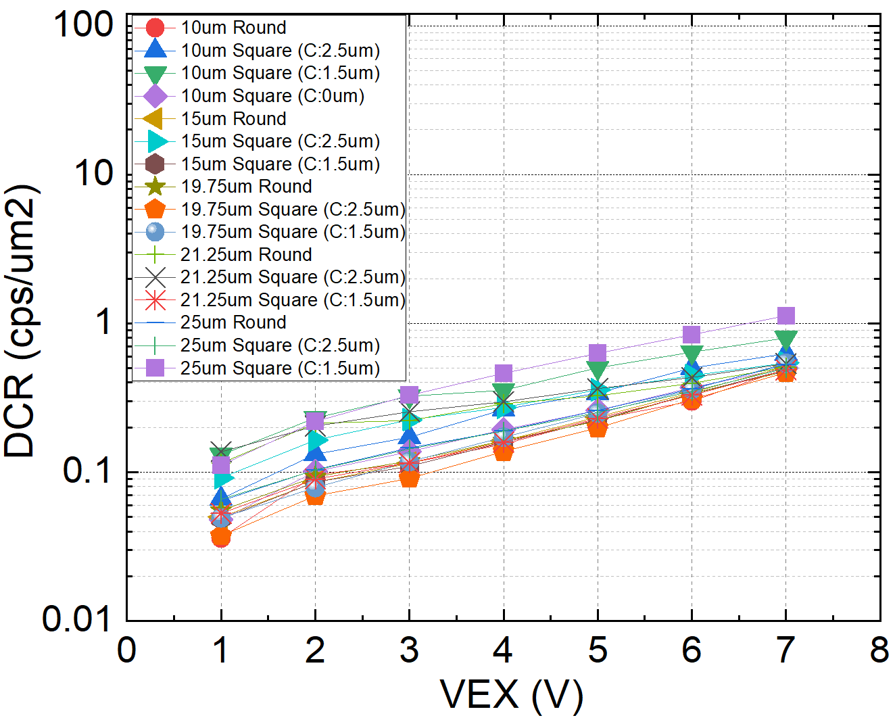}
    \caption{DCR versus excess bias voltage for the SPADs, comparing different active areas and curvature designs at room temperature.}
    \label{DCR_SPAD}
\end{figure}

Figure \ref{DCR_SPAD} presents the DCR of SPADs with varying curvatures and active area sizes as a function of excess bias. The tested samples have active area sizes ranging from 10~$\mu$m to 25~$\mu$m, with curvatures varying from a circular shape to a square design with a curvature radius of 1.5~$\mu$m. Notably, except for the most aggressive design, which has an active area of 25 µm and a curvature radius of 1.5 µm, the DCR normalized by area remains below 1 cps/µm² even at 7V excess bias for almost all SPAD shapes, demonstrating consistent performance across different geometries.

\begin{figure}[t]
    \centering
    \includegraphics[width=\linewidth]{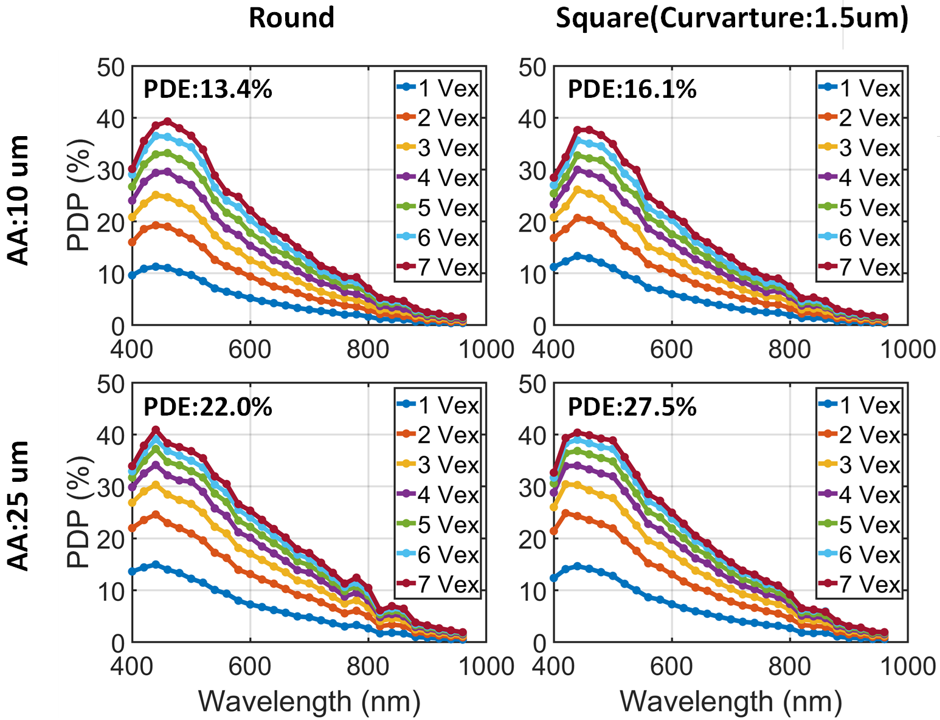}
    \caption{PDP spectrum of the SPAD comparing different active areas and curvature designs.}
    \label{PDP_SPAD}
\end{figure}

The measured PDP curves from 1~V to 7~V excess bias voltage and from 400~nm to 960~nm wavelengths are shown in Figure \ref{PDP_SPAD}. The PDE in the graph was calculated under the assumption that  4 \texttimes 4 SPAD array was designed with an Nwell-sharing technique.
Like the DCR measurements, the PDP also shows stable results regardless of the size of active area or curvature, recording a peak of around 40 \% at a wavelength of 460 nm.

These results demonstrate that layout optimization significantly increased the fill factor while maintaining a low DCR and avoiding PDP degradation. Specifically, In the scenario of 4\texttimes  4 SPAD array, the PDE increased from 13.4\% in the SPAD that has a 10 µm active area with a round shape to a maximum of 27.5\% in the SPAD with a 25 µm active area and a square shape.

\subsection{4\texttimes  4 SPAD macropixel operation}
\begin{figure}[t]
    \centering
    \includegraphics[width=1\linewidth]{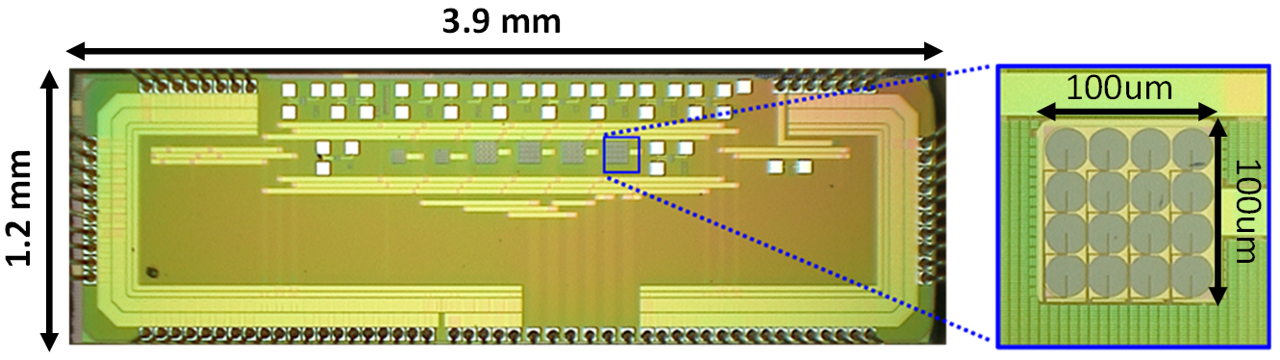}
    \caption{ Micrograph of the fabricated test chip. A region corresponding to one of the 4×4 SPAD macropixel variants is highlighted.}
    \label{Micrograph}
\end{figure}

\begin{figure}[t]
    \centering
    \includegraphics[width=1\linewidth]{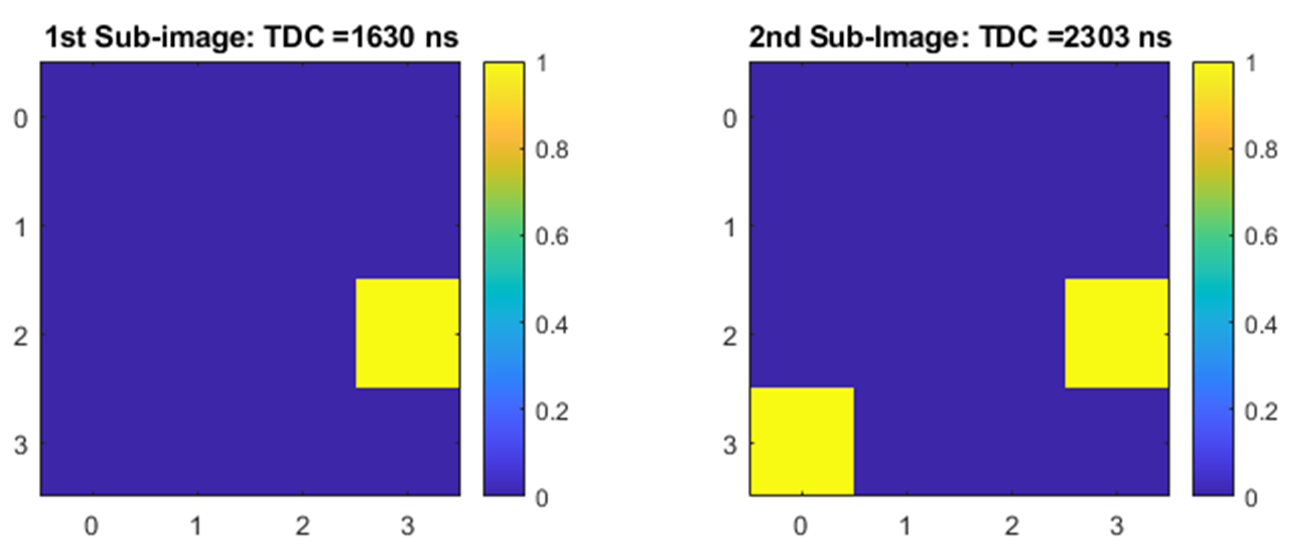}
    \caption{An example of the two sub-images captured by the 4×4 SPAD sensor macropixel module in a dark environment.}
    \label{Example}
\end{figure}

Figure \ref{Micrograph} presents the micrograph of the fabricated test chip, with a highlighted region corresponding to one of the 4×4 SPAD macropixel variants.  
An example of an acquired image in a dark environment is shown in Figure \ref{Example}, demonstrating the successful implementation of the two-sub-image acquisition, where each frame is captured with an associated timestamp, which is triggered by the first SPAD in the image. 

\begin{figure}[t]
    \centering
    \includegraphics[width=1\linewidth]{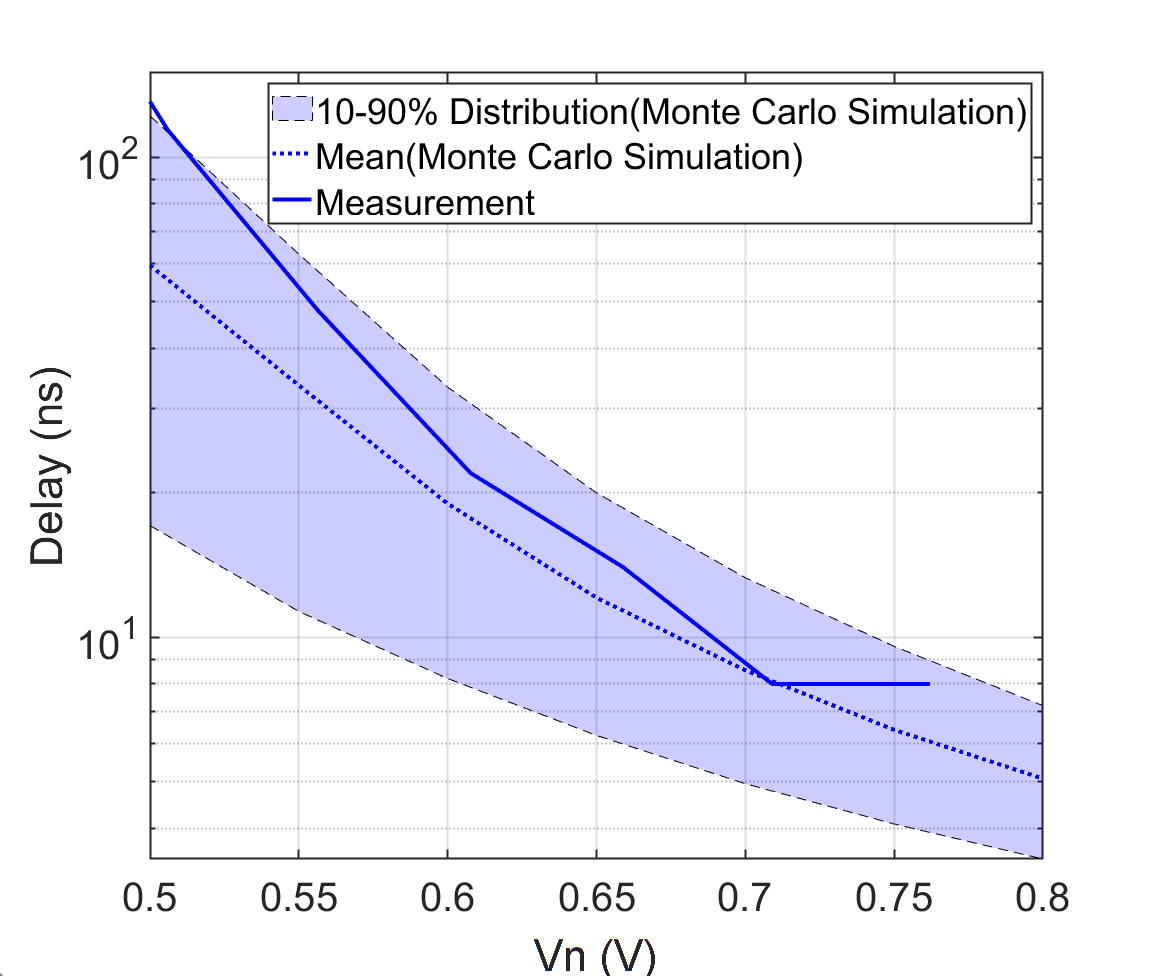}
    \caption{Range of the delay of the current starved inverter.}
    \label{Delay_CSI}
\end{figure}

Figure~\ref{Delay_CSI} shows the measured delay of the current-starved inverter as a function of the voltage applied to the NMOS current source, which is referred to as Vn Delay in Figure~\ref{System_Overview}. The solid blue line represents the measurement results obtained from the test structure, while the dashed blue line indicates the mean value extracted from Monte Carlo simulations. The 10–90\% variation range is depicted as a shaded blue region. At Vn Delay = 0.7 V, both the measured and simulated mean delays converge at 8 ns, with a variation spanning 3.3 ns to 20 ns. This range closely aligns with the decay time of a typical plastic scintillator, enabling the system to effectively capture scintillation photon emissions, while rejecting unwanted noise.

\section{Conclusion \& Future work}

In this work, we developed and characterized SPAD layouts with different geometries and a 4×4 SPAD macropixel architecture for future large-scale high-resolution particle sensors. Characterization results showed that the DCR remained around or below 1 cps/$\mu$m$^2$ at 7V excess bias, independent of SPAD geometry, while the PDP peaked at around 40 \% at a wavelength of 460 nm, demonstrating the PDE improvement without degrading noise performance. Additionally, the ability to capture two sub-images with timestamps and a tunable delay, which matches the decay time of plastic scintillators, was successfully demonstrated.  

Future work includes developing a large-scale SPAD sensor based on the optimized SPADs and 4×4 macropixel modules presented in this study. The final PlatonSPAD is currently under design, and its key specifications are summarized in Table~\ref{tab:PlatonSPAD_specs}. Further development will focus on intra- and inter-chip synchronization, crosstalk, and an efficient data acquisition system to enable large-scale detector applications.

\begin{table}[htbp]
    \centering
    \caption{Specifications for the final PlatonSPAD Sensor.}
    \label{tab:PlatonSPAD_specs}
    \renewcommand{\arraystretch}{1.1} 
    \setlength{\tabcolsep}{5.5pt} 
    \begin{tabular}{|p{3.5cm}|p{4cm}|} 
        \hline
        \textbf{Parameter} & \textbf{PlatonSPAD chip and system} \\
        \hline
        Chip Size & 9.0 × 9.0 mm$^2$ \\
                 & 192 × 272 SPADs  \\
                 & (48 × 68 macropixels) \\
        \hline
        Array Size & 8 × 8 chips \\
        \hline
        TDC Resolution & 200 ps \\
        \hline
        Pixel Pitch & 37.5 $\mu$m \\
        \hline
        Dark Count Rate (max) & $\sim$ 1 cps/$\mu$m$^2$ \\
        \hline
        Photon Detection Efficiency (PDE) &  10.8\% \\
        \hline
        Technology & CIS 110 nm \\
        \hline
    \end{tabular}
\end{table}
\sloppy
\section*{Acknowledgment}
This work was supported by the SNSF grant PCEFP2\_203261, Switzerland.

\bibliographystyle{IEEEtran}

\bibliography{references}
\end{document}